\documentclass[aps,pra,reprint,amssymb,amsmath,superscriptaddress,noeprint]{revtex4-2} 
\usepackage{graphicx}
\usepackage{bm}
\usepackage[colorlinks=true,citecolor=blue,linkcolor=blue,urlcolor=blue]{hyperref}
\usepackage{orcidlink}
\usepackage{physics}
\usepackage{layouts}

\newcommand{\mean}[1]{\langle #1 \rangle}

\usepackage{amsmath,amsthm,amssymb}
\usepackage{graphicx}
\usepackage[dvipsnames]{xcolor} 
\usepackage{color}

\usepackage{braket}

\begin{document}

\title{Benchmarking dynamical-structure-factor protocols in programmable neutral-atom geometries}

\author{Matteo Grotti\,\orcidlink{0009-0003-7245-1898}}
\email{matteo.grotti4@unibo.it}
\let\comma,
\affiliation{Dipartimento di Fisica e Astronomia dell'Universita' di Bologna, 40127 Bologna, Italy}
\affiliation{INFN, Sezione di Bologna, 40127 Bologna, Italy}

\author{Sergi Julià-Farré\,\orcidlink{0000-0003-4034-5786}}
\email{sergi.julia-farre@pasqal.com}
\let\comma,
\affiliation{PASQAL SAS, 24 rue Emile Baudot - 91120 Palaiseau,  Paris, France}
\author{Elisa Ercolessi\,\orcidlink{0000-0002-6801-5976}}
\email{elisa.ercolessi@unibo.it}
\let\comma,
\affiliation{Dipartimento di Fisica e Astronomia dell'Universita' di Bologna, 40127 Bologna, Italy}
\affiliation{INFN, Sezione di Bologna, 40127 Bologna, Italy}
\author{Alexandre Dauphin\,\orcidlink{0000-0003-4996-2561}}
\email{alexandre.dauphin@pasqal.com}
\let\comma,
\affiliation{PASQAL SAS, 24 rue Emile Baudot - 91120 Palaiseau,  Paris, France}

\begin{abstract}
    The dynamical structure factor is a central observable in condensed-matter physics, providing direct insight into the excitation spectrum and dynamical response of quantum materials. While it is traditionally accessed in real materials through inelastic neutron scattering, recent works have shown that analogous information can be extracted in quantum simulators using suitable dynamical protocols. So far, these approaches have been demonstrated mainly in paradigmatic integrable one-dimensional models. In this work, we use numerical emulations of a neutral-atom quantum processing unit to assess the feasibility of measuring the dynamical structure factor in a broader class of Ising-like spin systems. Beyond the standard one-dimensional transverse-field Ising chain, we benchmark the protocol in chains with dimerized interactions and in the two-dimensional transverse-field Ising model, where spectral properties are difficult to access with classical numerical methods at large scales. We further analyze the robustness of the protocol under realistic experimental conditions, including finite simulation times, pulse modulation, positional disorder, and laser noise. Our results show that neutral-atom quantum simulators can provide a practical route to probing dynamical response functions in regimes where classical simulations become increasingly demanding, paving the way toward experimental implementation.
\end{abstract}
\maketitle

\section{Introduction}
\label{sec:Introduction}

Over the past decades, quantum simulators~\cite{feynman_simulating_1982,manin_computable_1980,georgescu_quantum_2014} have emerged as a powerful alternative to classical numerical methods for investigating quantum many-body physics. This is particularly relevant for dynamical properties~\cite{andersen_thermalization_2025,bernien_probing_2017,chen_spectroscopy_2025,dag_emergent_2024,gonzalez-cuadra_observation_2024,keesling_quantum_2019,lamb_ising_2024, manovitz_quantum_2025,mark_observation_2025,wienand_emergence_2023,geimEngineeringQuantumCriticality2026,andersenPrecisionQuantumSimulation2026,sun_observation_2026,granetDynamicalStructureFactor2026}, which often remain difficult to access even with state-of-the-art approaches~\cite{vovrosh_simulating_2026} such as tensor networks or neural quantum states. A key observable for characterizing the dynamical response of quantum materials is the dynamical structure factor (DSF)~\cite{boothroyd_principles_2020}, routinely measured in magnetic materials through inelastic neutron scattering~\cite{rodriguez_macsnew_2008,granroth_sequoia_2010}. Recently, there has been a growing interest in accessing this quantity through local quench spectroscopy protocols in quantum simulators~\cite{knap_probing_2013,baez_dynamical_2020, villa_local_2020,piccinelli_circuit-differentiation_2026, buchs_validation_2026}, as demonstrated in paradigmatic one-dimensional integrable models using both universal digital devices~\cite{lee_benchmarking_2026,millarQuenchSpectroscopyMagnetic2026a} and analog platforms~\cite{bauer_progress_2025}. To make this approach broadly practical, it is crucial to move toward scenarios with richer DSFs, especially in nonintegrable models where classical calculations are possible only with sophisticated variational techniques~\cite{mendes-santos_highly_2023,drescher_dynamical_2023,ferrari_dynamical_2019}, whose accuracy and computational cost become increasingly challenging at large system sizes~\cite{baez_dynamical_2020}.
\begin{figure*}[t]
\centering
\includegraphics[width=\textwidth]{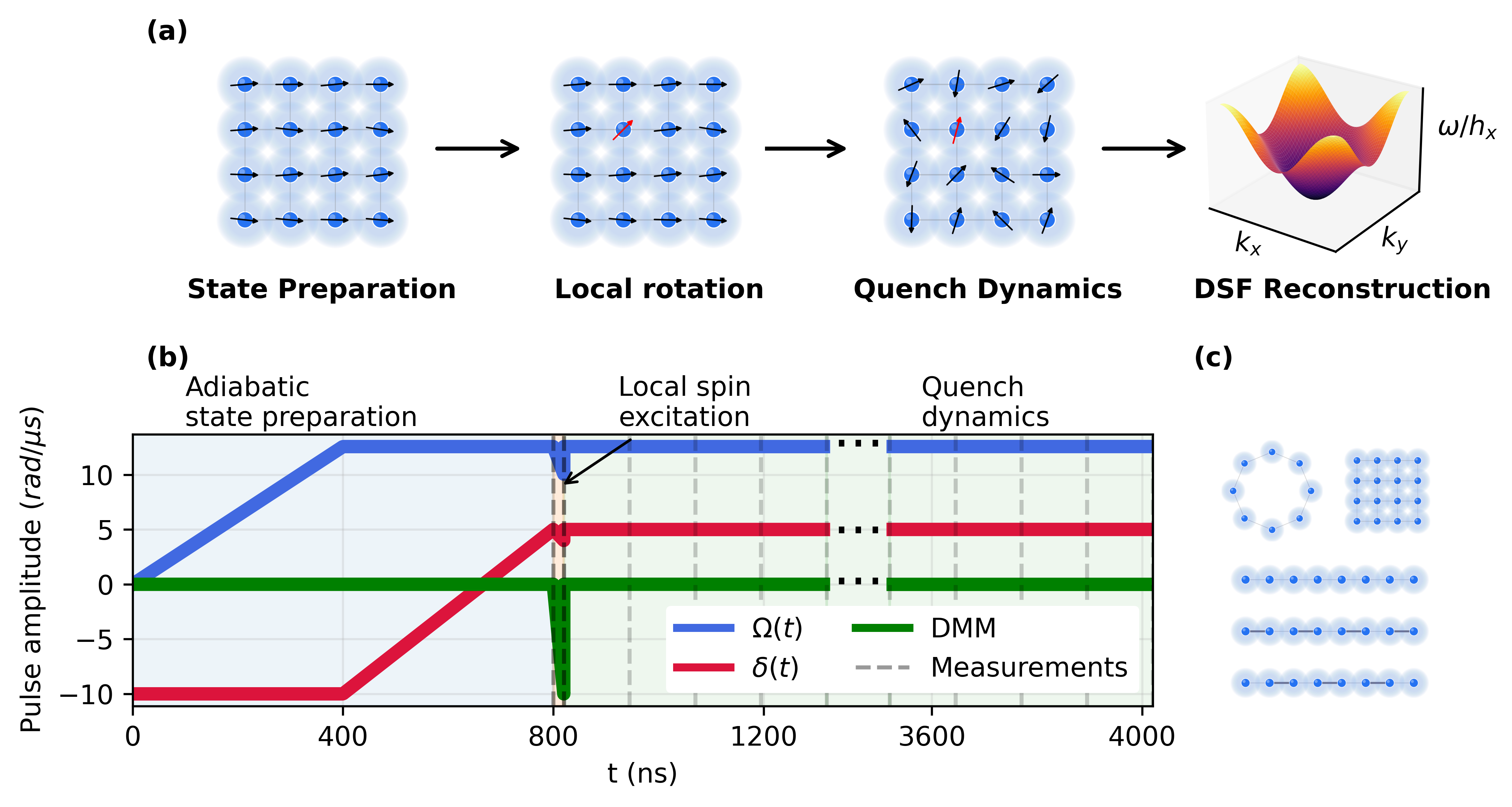}
\caption{(a) Sketch of the DSF measurement protocol adopted in this paper. (b) Sketch of the protocol sequence. (c) Sketch of atom registers that we consider in our paper: linear registers include Ising and SSH lattices (the latter both with and without weakly coupled edge sites) with open boundary condition while the remainder represents the Ising lattice with periodic boundaries.}
\label{fig:fig1}
\end{figure*}
In this work, we simulate numerically local quench spectroscopy protocols on a neutral-atom quantum processing unit (QPU)~\cite{saffman_quantum_2010,henriet_quantum_2020,browaeys_many-body_2020}. We show how these devices can provide relevant information about the excitation spectrum of a wide variety of spin models, ranging from several one-dimensional (1D) integrable examples to a two-dimensional (2D) case in which integrability is strongly broken. To this aim, we start by reviewing the ability of neutral-atom QPUs to simulate transverse field Ising models (TFIM) in a controllable and scalable setup~\cite{browaeys_many-body_2020}, and a dynamical protocol for measuring dynamical structure factors in this platform~\cite{knap_probing_2013,baez_dynamical_2020}. Then, we use state-vector and matrix-product state~\cite{schollwock_density-matrix_2011} emulators~\cite{silverio_pulser_2022, bidzhiev_efficient_2025} to benchmark such protocols in a plethora of different scenarios. First, we show how the protocol recovers the analytically predicted magnon bands of the 1D TFIM. Second, we show that richer spectrums can be accessed by the same protocol in the presence of 1D dimerized interactions, highlighting how different bulk and edge conditions are reflected in the spectrum of these systems. Third, we move to the DSF measurement protocol in the 2D TFIM, which we benchmark in small clusters. Finally, we also perform a numerical study of the impact of noise sources in the protocol for the 1D TFIM system, to assess the robustness of the neutral atom platform for extracting the DSF. 

\section{Model and methods}\label{sec:model}

\subsection{Transverse field Ising models with neutral atoms}
\label{subsec:model_ising}
Transverse field Ising models can be realized with $N$ neutral atoms in 2D programmable optical tweezer arrays. In this platform, a local qubit labeled by an index $i$ is encoded in the ground state $\ket{g}$ and a highly excited Rydberg state $\ket{r}$ of each trapped atom. The Hamiltonian governing the many-atom dynamics can be decomposed as~\cite{browaeys_many-body_2020}
\begin{equation}\label{eq:ham_qpu}
    \hat{H}_\textrm{QPU}=\hat{H}_\textrm{TFIM} + \hat{H}_l + \hat{H}_{\textrm{det}}.
\end{equation}
The first term is the target TFIM, that is 
\begin{equation}
\label{eq:tfim}
    \hat{H}_{\text{TFIM}} = \sum_{i<j} J_{ij} \hat{\sigma}^z_i\hat{\sigma}^z_j+
    h_x(t) \sum_i \hat{\sigma}^x_i.
\end{equation}
Here $\hat{\sigma}^\mu_i$ are the usual Pauli operators on site $i$. $J_{ij}=JR^6/r_{ij}^6$ is a short-range Van der Waals interaction between $(i,j)$ atom pairs separated by a distance $r_{ij}$, and $J$ and $R$ are defined as the nearest-neighbor interaction and distance, respectively. The transverse field is achieved via a two-photon laser transition with modulable Rabi frequency $\hbar\Omega(t)=2h_x$. For concreteness, in the numerical emulations of the QPU we fix the energy scale by setting a maximum $\Omega/(2\pi)=2 \textrm{MHz}$. The second term
\begin{equation}\label{eq:ham_local}
    \hat{H}_l =-\frac{\hbar\delta_l(t)}{2}\hat{\sigma}^z_l
\end{equation}
represents a single-site detuning term, which can be used to address the $l$ qubit of the array.
Finally, the last term 
\begin{equation}
    \hat{H}_{\textrm{det}} = \sum_i\left(\sum_{j}J_{ij} -\frac{\hbar\delta(t)}{2}\right)\hat{\sigma}^z_i
    \label{eq:det_compensation}
\end{equation}
originates from an interaction-induced level shift that is nearly uniform in regular arrays, and the global detuning $\delta(t)$ of the two-photon laser transition. The global detuning can be used to shift the energy of the initial state $\ket{g\dots g}$ in adiabatic state preparation protocols, but also to compensate the interaction shift for bulk sites. In this latter case, $\hat{H}_\textrm{QPU}\approx \hat{H}_\textrm{TFIM}$, leaving only residual inhomogeneities near edges (and the intentionally applied local $\hat{H}_l$).

\subsection{Dynamical structure factor protocol: state preparation, local rotation and many-atom dynamics}
\label{subsec:model_DSF_Green}
Given the general TFIM Hamiltonian defined by Eq.~\eqref{eq:tfim}, here we review the protocol that we will use to measure the DSF~\cite{knap_probing_2013,baez_dynamical_2020,bauer_progress_2025}. The main challenge is to measure the retarded Green's function in the z-basis of a low-energy state $\ket{\Psi_0}$
\begin{equation}
    G_{z,z}^\textrm{ret}(i,j,t)=-\frac{i}{2}\mean{{\hat{\sigma}}^z_i(t){\hat{\sigma}}^z_j(0)-{\hat{\sigma}}^z_j(0){\hat{\sigma}}^z_i(t)}_{\ket{\Psi_0}}.
\end{equation}
Note that in this work, for concreteness, we focus on paramagnetic regimes with $h_x/J\gg 1$. In this case, the characteristic magnetic excitations are magnons, well-captured by the z component of the DSF, $G_{z,z}^\textrm{ret}(i,j,t)$. As a first step, the low-energy state is prepared from the initial QPU state $\ket{g\dots g}$ via a standard quasi-adiabatic protocol of duration $T_\textrm{prep}=800\,\textrm{ns}$ sketched in Fig.~\ref{fig:fig1} (see App.~\ref{app:protocols} for details). This initial step leads to a low-energy state of $\hat{H}_\textrm{TFIM}$, parametrized by the ratio $J/h_x$~\footnote{Note that similar results could be obtained by  preparing a low-energy product state, such as $\ket{\Psi}_\textrm{prod}=(\ket{g}-\ket{r})^{\otimes N}$ for the case studied here. Compared to the finite-time limitation of our annealing protocol, here one needs to implement an analog gate to rotate the initial state $\ket{g\dots g}$ in the presence of the always-on Rydberg interactions.}.
Subsequently, the out-of-time correlators can be measured in the neutral atom platform using equal-time correlators in a Raman spectroscopy scheme, following the protocols introduced Refs.~\cite{knap_probing_2013,baez_dynamical_2020}. In a nutshell, one needs to compare the Hamiltonian evolution of $\ket{\Psi_0}$ and $\ket{\Phi_0}\equiv U^{(j)}\Psi_0$, where $U^{(j)}$ is a unitary $\pi/4$ rotation along the $z$-axis on a site $j$. To experimentally implement $U^{(j)}$, here we consider the local Hamiltonian of Eq.~\eqref{eq:ham_local}, that is
\begin{equation}
    \hat{U}^{(j)}=e^{-i\hat{H}_{\text{l}}T_\textrm{l}},
\end{equation}
with $\frac{\delta_lT_\textrm{l}}{2}=\pi/4$. We consider a fast rotation time of $T_\textrm{l}< 100$ ns, so that the effect of other terms in $\hat{H}_\textrm{QPU}$ can be neglected. The retarded Green's function is expressed as
\begin{equation}
\begin{split}\label{eq:model_retardedgreen}
    &G_{z,z}^\textrm{ret}(i,j,t) = \bra{\Phi_0^j(t)}\hat{\sigma}_i^z \ket{\Phi_0^j(t)} \\
    &- \frac{1}{2}\bra{\Psi_0(t)}\hat{\sigma}_i^z \ket{\Psi_0(t)}-\frac{1}{2}\bra{\Psi_0}{\hat{\sigma}}_j^z{\hat{\sigma}}^z_i(t){\hat{\sigma}}_j^z\ket{\Psi_0}.
\end{split}
\end{equation}
Importantly, the second and third terms above vanish for equilibrium states of transverse field Ising models of the form of Eq.~\eqref{eq:tfim}, e.g. $G_{z,z}^\textrm{ret}(i,j,t) \approx  \bra{\Phi_0^j(t)}\hat{\sigma}_i^z \ket{\Phi_0^j(t)}$. Note that, in the QPU protocol, this cancellation is only approximate because of finite preparation time and boundary-induced detuning inhomogeneities. We therefore include the correction associated with $\frac{1}{2}\bra{\Psi_0(t)}\hat{\sigma}_i^z \ket{\Psi_0(t)}$, while neglecting the last term in Eq.~\eqref{eq:model_retardedgreen}.
Finally, at zero temperature, i.e., when $\ket{\Psi_0}$ is the ground state of the system, the DSF is simply $S^{zz}(k,\omega)=-\frac{1}{\pi}\textrm{Im}G_{z,z}^\textrm{ret}(k,\omega)$. In order to perform the time and space Fourier transforms, we consider a time evolution of $T=3200\,\textrm{ns}$ after the local rotation, under a constant QPU Hamiltonian approximately equal to $\hat{H}_\textrm{TFIM}$ in Eq.~\eqref{eq:tfim}, parametrized by $J/h_x$. Note that during this stage we minimize the impact of $\hat{H}_\textrm{det}$ by approximately cancelling the contributions from interactions with the global detuning $\delta(t)$, see details in Appendix~\ref{app:protocols}.  Furthermore, note that in general we need to simulate $N+1$ dynamical evolutions, being $N$ the number of qubits of the system: the quench dynamics has to be run after flipping each possible qubit ($N$ runs in total) and an extra run with no single qubit flip, needed to compute the offset of qubits magnetization (second term of the right-hand side of Eq. \eqref{eq:model_retardedgreen}). For the finite lattices considered in this work, we apply the local rotation to each of the $N$ sites in order to improve the reconstruction of the momentum-resolved response. In larger bulk lattices accessible on a real QPU, a single local rotation applied to a bulk site is expected to suffice when translational invariance can be assumed. We also adopt this simplification for the PBC systems studied here, where translational invariance is exact. Further details on the computation of the Fourier Transform of the retarded Green's function can be found in App. \ref{app:FT}.\\
To perform the numerical emulations of the QPU protocols presented in the next sections, we use the \texttt{pulser} library~\cite{silverio_pulser_2022} and its backends~\cite{bidzhiev_efficient_2025} based on state-vector (\texttt{emu-sv}) and matrix product states (\texttt{emu-mps}). More details on the numerical methods are provided in Appendix~\ref{app:numerics}.

\section{One-dimensional Ising model}
\begin{figure}[t]
    \includegraphics[width=\columnwidth]{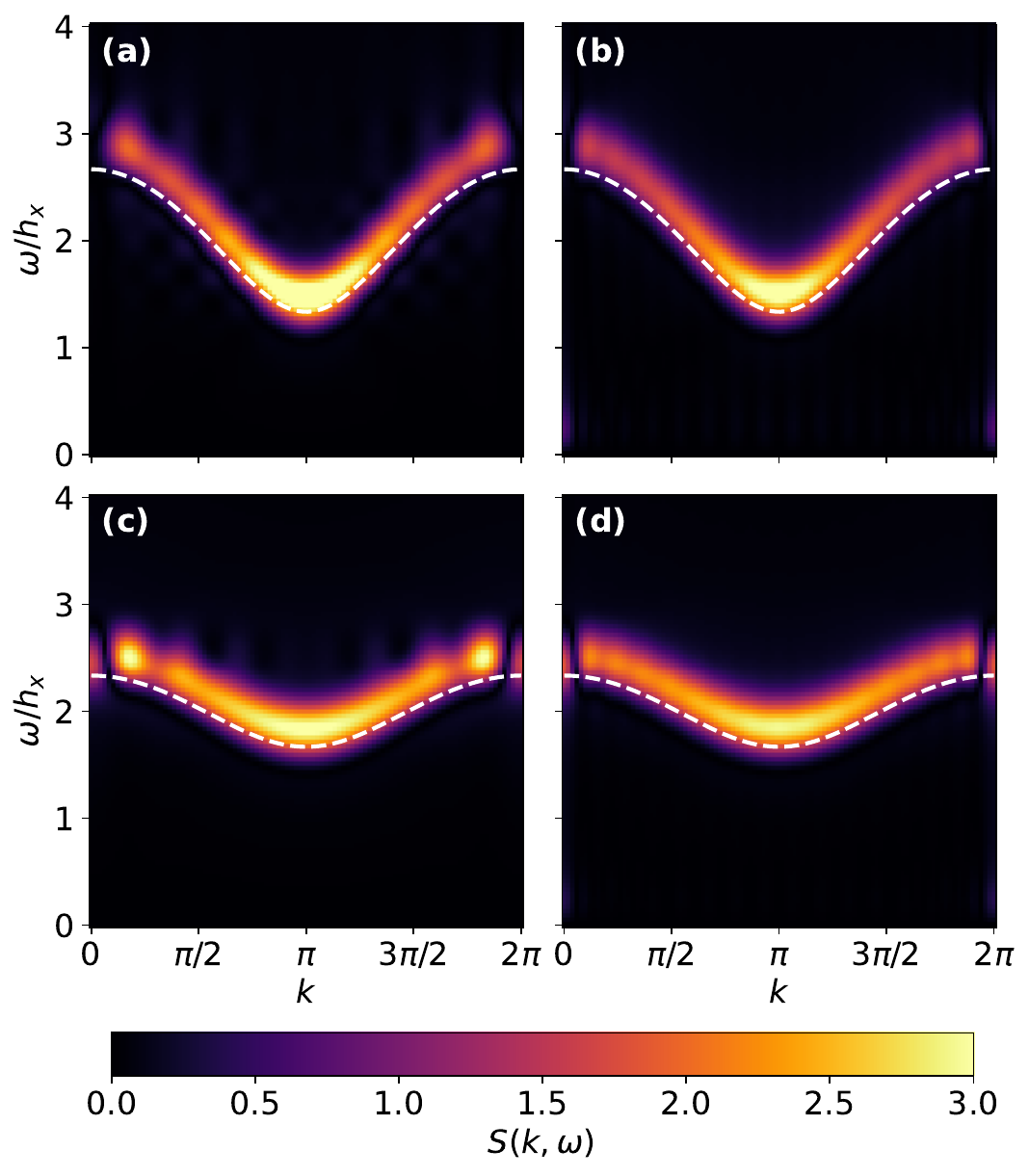}
    \caption{Dynamical structure factor of the 1D Ising model ($L=16$) obtained in a neutral-atom QPU protocol (ideal state-vector emulation). We consider different values of $J/h_x$ and different boundary conditions: $J/h_x=1/3$ with PBC (a), OBC (b) and $J/h_x=1/6$ with PBC (c) or OBC (d). The white dashed line represents the theoretical excitation band.}
    \label{fig:1Dnondim}
\end{figure}
We start by showing how the toolbox outlined in the previous section can be used to extract the DSF in the paradigmatic 1D TFIM case of Eq.~\eqref{eq:tfim}, achieved by placing a neutral atom array in a ring geometry (see Fig.~\ref{fig:fig1}d). We note that the DSF measurement in this simple case has already been studied in the literature, both theoretically~\cite{baez_dynamical_2020, li_detecting_2022} and experimentally~\cite{bauer_progress_2025}. Since this model can be solved analytically neglecting beyond-NN interactions, the predicted magnon bands (see App.~\ref{app:dsf} for details) are useful to benchmark the neutral atom protocol. Figure \ref{fig:1Dnondim} shows the dynamical structure factor for a one-dimensional Ising model with $N=L=16$ atoms, obtained via a state-vector emulation of the QPU protocol described in the previous section. In particular, we show the results at $h_x/J=3$ with periodic (a) and open (b) boundary conditions, and $h_x/J=6$ with periodic (c) and open (d) boundary conditions. The excitation band in both cases fits the theoretical magnon band, with a larger occupation at lower energy, corresponding to $k=\pi$. Furthermore, the bands exhibit the expected dependence with decreasing $h_x/J$: the almost classical flat band at $h_x \gg J$ becomes more dispersive when approaching the quantum critical point, predicted at $h_x/J=1$ for the NN model in the thermodynamical limit. The comparison against the analytical bands also reveal that, even in the absence of explicit noise sources, small deviations from the ideal analytical dispersion remain visible. These originate from several intrinsic limitations of the protocol, including the finite duration of the state-preparation sequence, the finite size and evolution time, and boundary effects associated with the $\hat{H}_{\textrm{det}}$ present in the QPU Hamiltonian of Eq.~\eqref{eq:ham_qpu}, which breaks the exact integrability of the system. The latter become apparent when comparing open and periodic boundary conditions. Interestingly, the deviations are most pronounced at low momenta, indicating an increased sensitivity of long-wavelength excitations to perturbations that vary weakly across the lattice.

\section{One-dimensional SSH-Ising model}

\begin{figure}[t]
    \includegraphics[width=\columnwidth]{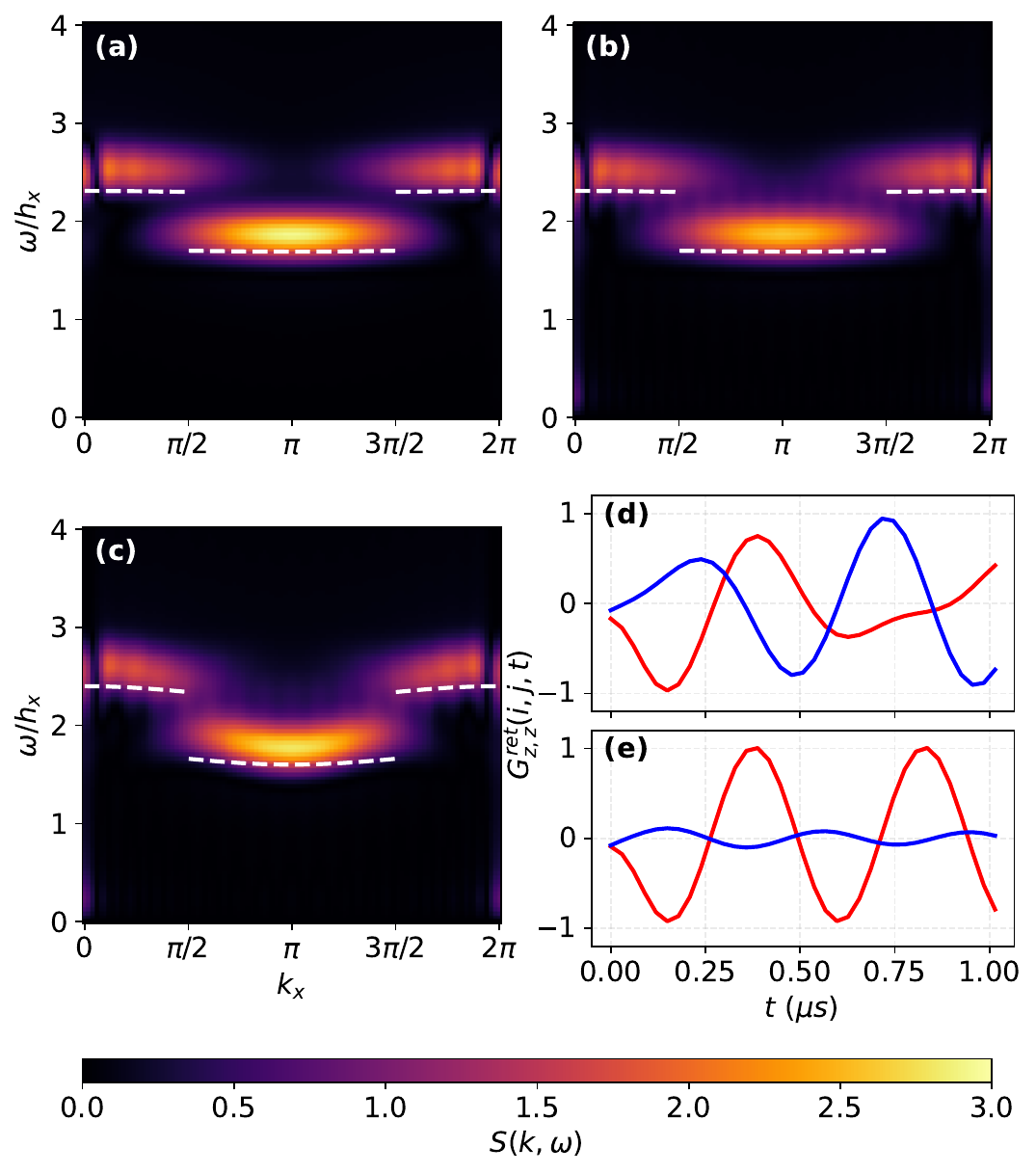}
    \caption{Dynamical structure factor of the 1D SSH-Ising model ($L=20$) obtained in a neutral-atom QPU protocol (ideal state-vector emulation). The system is dimerised with $J$ being the coupling between qubits belonging to the same dimer and $J'$ the coupling between neighbor qubits belonging to different dimers. In panel (a) $J'/h_x=0.01$, $J/h_x=0.3$ and the dimerization is chosen such that no edge qubit is left at the border while in panel (b) edge qubits are left at the border, being both $J'/h_x$ and $J/h_x$ the same as in panel (a). This choice of the interaction-to-field ratio leads to a flat band. Panel (c) shows the band when $J'/h_x=0.066$ and $J/h_x=0.333$, leaving edge qubits at the border. White dashed lines in panels (a)-(c) represent the theoretical excitation bands for the three cases. Panel (d) and (e) show the Green function in Eq.~\eqref{eq:model_retardedgreen} (third term is neglected) over time for the flipped qubit (red) and its nearest neighbor (blue) for a dimerised lattice with $J'/h_x=0.01$ and $J/h_x=0.3$ without leaving edge states and leaving edge states, respectively.}
    \label{fig:1Ddim}
\end{figure}
We now turn to a generalized 1D TFIM, in which alternating interaction strengths, achieved by dimerized chain geometries (see Fig.~\ref{fig:fig1}(d)), enrich the physics of the model. The system in Eq.~\eqref{eq:tfim} takes the form of a Su-Schrieffer-Hegger (SSH) spin Hamiltonian, well-known as a paradigmatic example of a topological quantum system in 1D~\cite{su_solitons_1979}. This case is also analytically solvable neglecting couplings beyond consecutive atoms, with a predicted band opening at $k=\pi/2, 3\pi/2$ associated to the doubling of the lattice unit cell, see App.~\ref{app:dsf}. This example highlights the flexibility of Rydberg-atom arrays, whose programmable geometry allows the implementation of engineered lattice structures with non-trivial excitation spectra. The DSF results obtained with the state-vector QPU emulator of the neutral-atom protocol are shown in Fig.~\ref{fig:1Ddim}(a-c), for different dimerized registers. A first remark is that, in all cases, we observe a good matching with the two-band picture predicted analytically (white dashed lines). Furthermore, for fixed Hamiltonian parameters $J'/h_x=0.01$ and $J/h_x=0.3$ it is also interesting to compare the case in which the dimerized chain hosts isolated edge sites, Fig.~\ref{fig:1Ddim}(a), and the case without, Fig.~\ref{fig:1Ddim}(b). When edge sites are present, the emergence of localized excitations is expected, due to their suppressed interaction with the rest of the chain. In the $\hat{H}_\textrm{TFIM}$, theory predicts them to be topologically protected at the energy in the middle of the gap, $\omega=2h_x$, under local perturbations preserving chiral symmetry. In the present case, the chiral symmetry is broken both by the presence of the $r^{-6}$ interaction tail in $\hat{H}_\textrm{TFIM}$, as well as the extra term edge term $\hat{H}_\textrm{det}$. While the delocalization in momentum space, and the small relative weight compared to the bulk states, makes it difficult to distinguish such edge-localized modes in Fig.~\ref{fig:1Ddim}(b), a better probe is given by the site-resolved response of the system, shown in Fig.~\ref{fig:1Ddim}(d-e). Figure~\ref{fig:1Ddim}(d) shows that, when the initial rotation is applied to an edge site that is coupled to the bulk via the stronger interaction $J'$, the local perturbation propagates to the rest of the chain, and we do not observe a well-defined local frequency. In contrast, when the initial rotation acts on an edge site that is only weakly coupled to the chain by a small $J'$, see Fig.~\ref{fig:1Ddim}(e), its magnetization is excited with a well-defined local frequency. The latter can be understood from an effective single-spin picture. Since the global detuning compensates the interaction shift of bulk sites, an edge site experiences a residual longitudinal field due to the missing neighboring interaction. This residual field has magnitude $J$, leading to a local two-level splitting
$\omega_{\rm edge}\simeq 2\sqrt{h_x^2+J^2}$.  Finally, Fig.~\ref{fig:1Ddim}(c) highlights that, as in the non-dimerized case, the band becomes more dispersive when increasing interactions. In particular, this effect is controlled by the smaller interaction $J'/h_x$.

\section{Two-dimensional Ising model}
\begin{figure}[t]
    \includegraphics[width=\columnwidth]{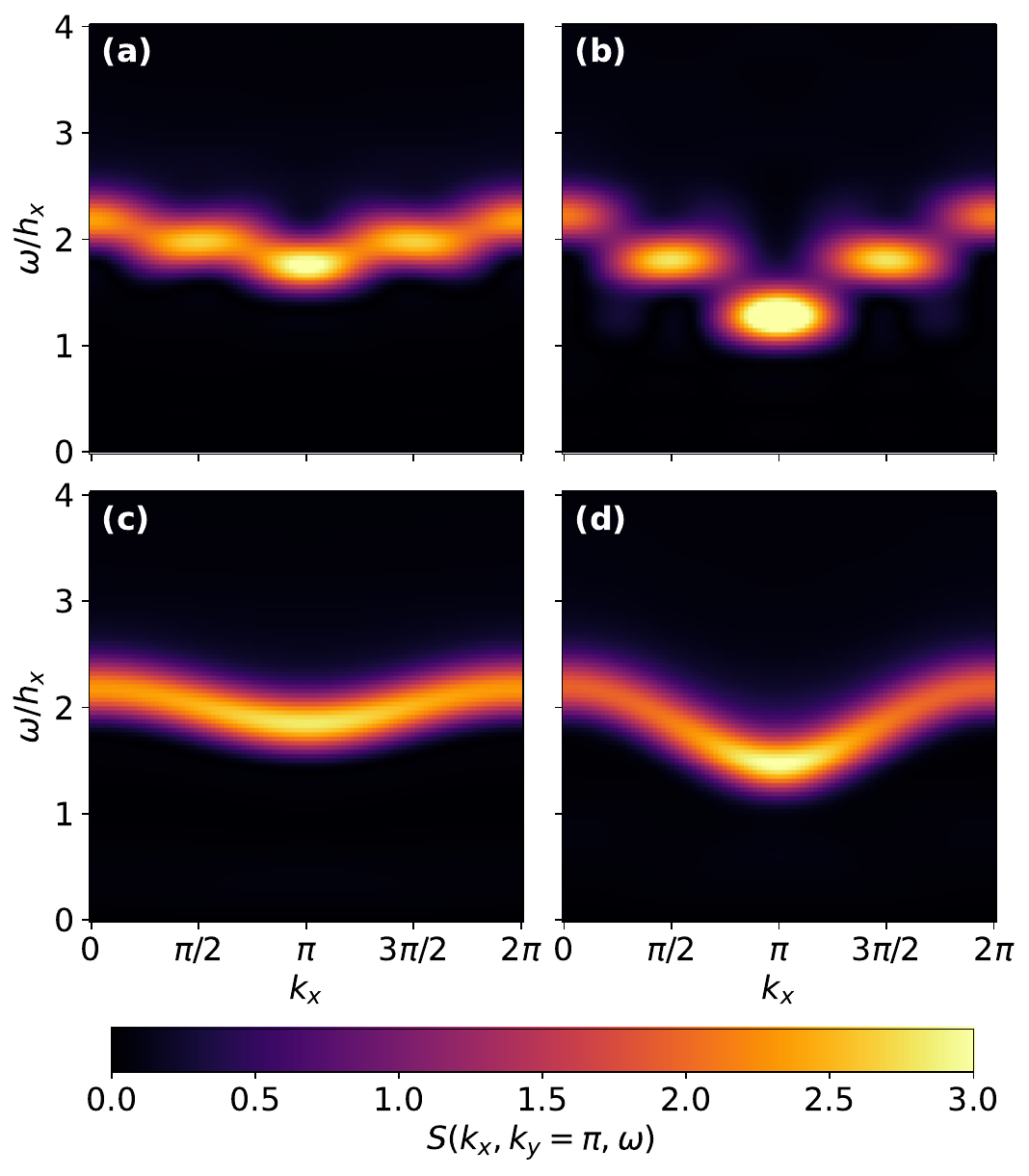}
    \caption{Dynamical structure factor of the 2D TFIM obtained in a neutral-atom QPU protocol with different values of $J/h_x$ and different boundary conditions: the system is either a $4\times 4$ lattice with PBC and $J/h_x=0.1$ (a) or $J/h_x=0.2$ (b), while it is a $6\times 6$ lattice with OBC and $J/h_x=0.1$ (c) or $J/h_x=0.2$ (d). The dynamics of $4\times 4$ lattices has been emulated through \texttt{emu-sv} state-vector emulator while the dynamics of $6\times 6$ lattices has been emulated through \texttt{emu-mps} with bond dimension $\chi=512$.}
    \label{fig:2D}
\end{figure}
Let us now discuss the 2D case. Compared to the 1D cases studied in the previous sections, the nonintegrable nature of the 2D TFIM prevents an analytical treatment even in the NN model. Nevertheless, important information about its DSF has been obtained using specialized classical numerical approaches~\cite{hamer_critical_2006, hamer_one-particle_2006, mendes-santos_highly_2023}, which however become increasingly demanding in system size and evolution times. Here we limit our QPU emulator benchmark to small systems and away from the quantum critical point, to assess the signal expected in larger arrays accessible to a real QPU. First, we consider a small $N=4\times 4=16$ square lattice, that still can be solved with the QPU state-vector emulator. Since the edge effects due to $\hat{H}_{\textrm{det}}$ are largely enhanced for such small cluster, we impose artificial PBC in the interactions of the atomic array. Figs.~\ref{fig:2D}(a-b) displays the resulting dynamical structure factor for the 2D TFIM with (a) $h_x/J=10$ and with (b) $h_x/J=5$. We focus on a fixed cut $k_y=\pi$ to suppress protocol-induced contributions more present at $\mathbf{k}=(0,0)$. Similar to the 1D case, we observe an excitation band around $\omega/h_x\approx 2$, which becomes more dispersive for decreasing $h_x/J$. Here it is also clear that finite-size effects lead to a poor resolution in the 2D k-space. To increase system size, we now move to the $N=6\times 6=36$ case, which we solve with realistic OBC using MPS methods with a considerably large bond dimension $\chi=512$. The results, shown in Fig.~\ref{fig:2D}(c-d) for $h_x/J=10$ and $h_x/J=5$, respectively, are in line with the $N=16$ case, but we observe a substantial improvement in the momentum resolution, leading to the emergence of more continuous and well-defined excitation bands. While here the MPS method remains accurate (see App.~\ref{app:bond_dimension}) since we investigate the DSF relatively deep in the PM phase, systems of this size already approach the limits of classical numerical methods. In particular, their accuracy is expected to worsen close to the quantum phase transition to the interaction-dominated antiferromagnetic phase, expected to be at $h_c/J\lesssim 3$ for the $\hat{H}_\textrm{TFIM}$ with short-range interactions (more details can be found in App.~\ref{app:bond_dimension}.
This highlights the potential of running the QPU protocol as a tool for exploring the DSF in larger two-dimensional quantum systems beyond the reach of exact classical methods.

\section{Impact of noise sources}
\begin{figure}[b]
    \centering
        \includegraphics[width=\columnwidth]{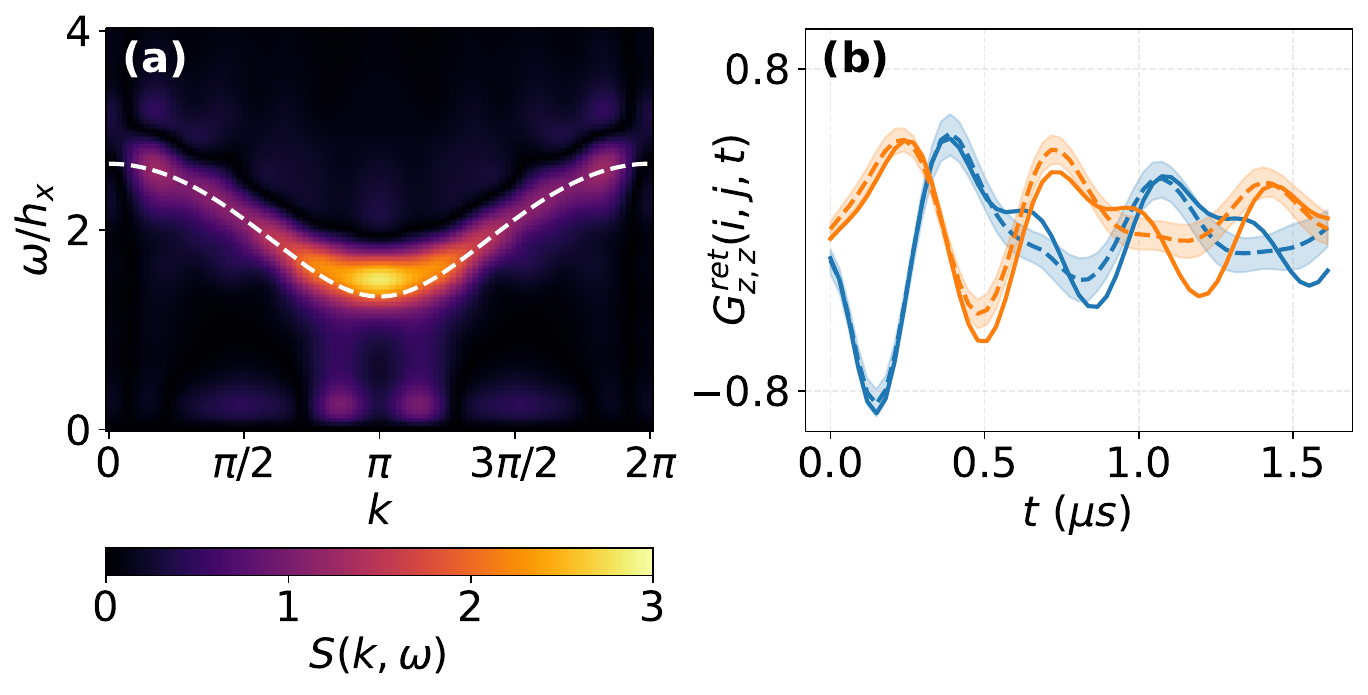}
    \caption{(a) Dynamical structure factor of 1D Ising model obtained in a neutral-atom QPU protocol (ideal state-vector emulation) with PBC and $J/h_x=0.333$ with noisy algorithm. The white dashed line represents the theoretical magnon band. (b) Green function of Eq.~\eqref{eq:model_retardedgreen} (third term is neglected) as a function of time of the flipped qubit ($j=i$, blue) and its nearest neighbour ($j=i+1$, orange). Solid lines refer to the noiseless case while dashed lines refer to the noisy one. Shaded areas represent the uncertainty on the average of the noisy trajectories computed with bootstrap method (App.~\ref{app:noise_model}).}
    \label{fig:noisy}
\end{figure}

We also investigate the robustness of the protocol against noise by emulating the effect of several noise sources (see Appendix \ref{app:noise_model} for more details) that are representative of current experimental platforms~\cite{de_leseleuc_analysis_2018,silverio_pulser_2022}. For concreteness, here we focus on the 1D TFIM. Although analogous noisy simulations could in principle be performed for the 2D system sizes considered in the previous section, they would substantially increase the computational cost. The 1D setting therefore provides a controlled benchmark to isolate the impact of stochastic noise sources. The influence of noise is first visible in the real-space magnetization dynamics shown in Fig.~\ref{fig:noisy}(b), where fluctuations and decoherence progressively distort the ideal evolution. After transformation to momentum-frequency space, these imperfections manifest as a broadening and increased background noise in the reconstructed magnon band, as shown in Fig.~\ref{fig:noisy}(a). Nevertheless, the dominant spectral features remain clearly identifiable, and the overall dispersion retains the qualitative characteristics predicted by the analytical theory. These results indicate that the protocol is resilient to realistic levels of experimental noise and remains capable of extracting the essential properties of the excitation spectrum.\\
Additional sources of imperfections, not explicitly included in the present analysis, may further affect the reconstruction accuracy. These include errors in the implementation of the initial $\pi/4$ rotation, due to deviations in the short-pulse time duration and local detuning value, and residual calibration inaccuracies in laser and register parameters. Furthermore, in Fig. \ref{fig:noisy}(a) we plot the Fourier transform of the retarded Green function obtained from the mean over 50 noisy trajectories. Since the averaging is performed before the Fourier transform, this procedure smooths out the fluctuations between individual trajectories and therefore yields a more coherent signal than would be expected from an actual experimental implementation, where independent measurements would sample different noisy trajectories. The latter would additionally retain the statistical fluctuations associated with the dispersion of the individual data points. Besides noise sources altering the unitary dynamics governed by Eq.~\eqref{eq:ham_qpu}, the requirements for the adiabatic state preparation protocol are known to become more stringent in larger arrays, specially close to the quantum phase transition ($h_x/J\approx 1$ in 1D, and $h_x/J\approx 3$ in 2D). 

\section{Conclusions and outlook}
In this work, we have emulated neutral-atom pulse protocols to extract the dynamical structure factor of transverse field Ising models in different geometries. First, we have investigated the dynamical structure factor of the 1D chain, showing how the experimental protocol recovers the expected magnon band behavior. To go beyond this paradigmatic 1D case, we have also shown how a richer 1D chain with dimerized spacings leads to a more complex dynamical structure factor with a gap separating an upper and a lower band. In 2D, we have performed a small-cluster study, at $N=4\times 4$ and $6\times 6$, to benchmark the expected 2D signal, and highlight the numerical hardness of simulating the protocol numerically. Finally, we have performed a noise analysis for the 1D case, which showed that the scheme is expected to be robust against the typical noise levels of current neutral-atom devices.
This proposal thus paves the road for the experimental implementation. In particular, the dynamical structure factor of 2D Ising models is directly relevant to real materials~\cite{leclerc_one--one_2026}. Looking ahead, it would be interesting to extend our results to other spin Hamiltonians, such as $XY$ or Heisenberg models, which can be engineered in neutral-atoms~\cite{scholl_microwave_2022,nishad_quantum_2023}.
\acknowledgments
Pasqal acknowledges funding from the European Union under the projects PASQuanS2.1 (HORIZON-CL4-2022-QUANTUM02-SGA, Grant Agreement 101113690).
\newpage

\appendix
\section{Adiabatic state preparation protocol and Hamiltonian evolution}\label{app:protocols}

The preparation of the low energy initial state of the quench is achieved through a pseudo-adiabatic protocol. Its duration is set to $T=800\;ns$ even though $T$ is an arbitrary hyperparameter of the full DSF measurement protocol (depicted in Fig.~\ref{fig:fig1}). The adiabatic pulse is made out of two separate pulses, each lasting a time $T/2$. \\
\begin{itemize}
    \item Since the initial state $|g\rangle^{\otimes N}$ is the ground state of $\hat{H}=\frac{\delta}{2}\sum_i\sigma^z_i$ with $\delta<0$, the initial values of Rabi and detuning frequencies are set to $\Omega=0$ and $\delta<0$ (we choose $\delta=-10\; \textrm{rad}/\mu s$;
    \item The former pulse raises the Rabi frequency up to its maximum value equal to $\Omega=12.56\; \textrm{rad}/\mu s$, leaving the detuning constant;
    \item The latter pulse is used to bring the detuning to the value needed to compensate for the interaction shift for bulk sites as described in Sec.~\ref{subsec:model_ising}. Consider Eq. (\eqref{eq:det_compensation}): to compute $\delta$, we set $\hat{H}_{det}=0$ and we get $\delta_{i_{bulk}}=\frac{2}{\hbar}\sum_jJ_{i_{bulk}j}$. In this way, the current Hamiltonian at the end of the state preparation sequence is exactly the TFIM Hamiltonian of Eq. (\eqref{eq:tfim}). 
\end{itemize}

\begin{figure}
    \centering
    \includegraphics[width=0.9\columnwidth]{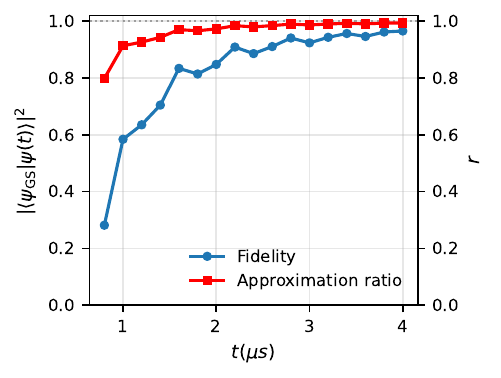}
    \caption{Approximation ratio (red) and fidelity (blue) of the atom state after the adiabatic state preparation protocol.}
    \label{fig:state_prep_assessment}
\end{figure}

To assess the effectiveness of the state-preparation protocol employed in the computation of the dynamical structure factor (Sec. \ref{subsec:model_DSF_Green}), we evaluate both the approximation ratio, defined as $r = E_{\mathrm{fin}}/E_0$, where $E_{\mathrm{fin}}$ is the energy of the prepared state and $E_0$ is the ground-state energy of the target Ising Hamiltonian, and the fidelity, $F = |\langle \psi_{\mathrm{fin}} | \psi_{\mathrm{GS}} \rangle|^2$. These quantities are computed for different state preparation duration $t$ on a 1D Ising chain with $J/h_x = 0.333$, whose dynamical structure factor is shown in Fig. \ref{fig:1Dnondim}. The results are reported in Fig. \ref{fig:state_prep_assessment}.

As expected, both the approximation ratio and the fidelity rapidly approach unity as the preparation time increases. In the protocol adopted throughout this work, the preparation time is fixed to $t = 0.8,\mu\mathrm{s}$, for which we obtain $r \simeq 0.80$ and $F \simeq 0.28$. Although these values may appear relatively modest, the primary objective of the state-preparation stage is not to accurately reproduce the ground state, but rather to significantly reduce the energy of the system so that the subsequent single-qubit excitation becomes relevant for the computation of the dynamical structure factor. The DSF is remarkably robust with respect to the choice of the initial state for the quench dynamics. In fact, we could recover the excitation band for all possible system parameters we have taken into account despite the imperfect state preparation. 

For comparison, the energy per site of the initial product state $|\psi_{\mathrm{init}}\rangle = |0\rangle^{\otimes L}$ is $E_{|\psi_{\mathrm{init}}\rangle} \simeq 2.09$, corresponding to an approximation ratio of $r_{\mathrm{init}} \simeq -0.33$. Therefore, the pseudo-adiabatic preparation stage significantly lowers the energy of the system, achieving the declared goal of producing a substantially improved starting point for the subsequent quench dynamics.
\section{Numerical methods}\label{app:numerics}
The registers and pulse sequences used in this work were designed using the \texttt{pulser} library~\cite{silverio_pulser_2022}, which includes realistic pulse modeling with a finite bandwidth in the sequences described in the previous App.\ref{app:protocols}. In the following, we describe the different \texttt{pulser} backends~\cite{bidzhiev_efficient_2025} that we use to emulate the many-body dynamics. 

\subsection{Emulation of 1D chains with and without noise}\label{app:noise_model}
For simulating the ideal dynamics, i.e., without noise effects, of the 1D systems of $L=16$ we use the exact state-vector emulator \texttt{emu-sv}. The noisy simulations of Fig.~\ref{fig:noisy}, in 1D at $L=16$ were done with the matrix-product-state emulator \texttt{emu-mps}, albeit with a bond dimension $\chi=256$ allowing the tensor-network ansatz to be exact at this small system size. We use a time discretization of $dt=5$ ns in the time-dependent variational principle. Table \ref{tab:noise_model} lists the noise sources we have taken into account for the noisy simulation ~\cite{de_leseleuc_analysis_2018,silverio_pulser_2022}, following the standard conventions in the \texttt{pulser} library. We can divide noise sources into four categories: 
\begin{table*}[t]
    \centering
\renewcommand{\arraystretch}{1.2}
    \begin{tabular}{|c|c|p{0.545\textwidth}|c|}
        \hline
        \textbf{Noise parameter} & \textbf{Name} & \textbf{Description} & \textbf{Value}  \\
        \hline\hline 
        $\sigma_{\Omega}/\Omega$ & Laser amplitude noise 
        & Standard deviation of a Gaussian distribution modeling relative shot-to-shot fluctuations of the Rabi frequency $\Omega$.
        & $0.01$ \\ 
        \hline
        $\sigma_{\delta}$ & Laser detuning noise 
        & Standard deviation of Gaussian fluctuations in the laser detuning $\delta$ across different experimental shots 
        & $0.05\textrm{rad}/\mu\textrm{s}$ \\ 
        \hline
        $\sigma_{\phi}$ & Phase noise 
        & Laser phase fluctuations leading to coherent dephasing in the driven Rydberg dynamics 
        & $^{*}$ \\
        \hline\hline
        $T$ & Temperature 
        & Effective atomic temperature contributing to Doppler-induced dephasing and motional effects 
        & $20\,\mu K$\\
        \hline
        $d_t$ & Trap depth 
        & Depth of the optical trapping potential defining atomic confinement strength in the tweezer array 
        & $70\,\mu K$ \\
        \hline
        $w_t$ & Trap waist 
        & Spatial width of the optical tweezer potential governing confinement geometry of individual atoms 
        & $0.84\,\mu m$ \\
        \hline
        \hline
        $\gamma_2$ & Dephasing rate 
        & Lindblad dephasing rate modelling loss of coherence in the computational basis 
        & $0.05\mu\textrm{s}^{-1}$\\ 
        \hline
        $\gamma_1$ & Relaxation rate 
        & Spontaneous decay rate from the Rydberg state to the ground state manifold 
        & $0.01\mu\textrm{s}^{-1}$\\ 
        \hline\hline
        $\eta$ & State preparation error 
        & Probability of imperfect initialization of atoms in the desired initial quantum state 
        & $0.018$\\ 
        \hline
    \end{tabular}
    \caption{Noise model considered in the emulation of the Pasqal neutral-atom quantum processing unit using \texttt{pulser}. The noise sources are grouped into laser imperfections, register noise, decoherence beyond the qubit manifold, and state preparation errors. $^{*}$Phase noise is constructed from the power spectral density of the lasers involved in the two-photon transition from ground to Rydberg state~\cite{de_leseleuc_analysis_2018}.}
    \label{tab:noise_model}
\end{table*}
\begin{itemize}
    \item Laser noise: includes fluctuations of $\Omega$ and $\delta$;
    \item Register noise: includes fluctuations of atom positions from shot-to-shot, and single-atom Doppler effect changing its local detuning;
    \item Environmental noise: includes effective dephasing and relaxation channels, mainly due to the finite lifetime of the $\ket{g}$ state and the off-resonant coupling to intermediate state used in the two-photon scheme to excite $\ket{r}$ from $\ket{g}$;
    \item SPAM noise: includes state preparation and measurement errors. Here we only include state preparation error because measurement errors can be pr-calibrated and corrected when reconstructing local observables from QPU bitstrings.
\end{itemize}
To perform an average of the noise model,  we simulated 50 Monte Carlo wavefunction trajectories, stochastically sampling the fluctuations described in App.~\ref{app:noise_model}. Effective channels beyond the qubit manifold are also taken into account in \texttt{emu-mps} via a quantum jump formalism. The associated uncertainty in local observables (shown in Fig.~\ref{fig:noisy}(b)) is then estimated by taking 15-85th trajectory percentiles.

\subsection{Emulation of the 2D square lattice}\label{app:bond_dimension}
\begin{figure}[h]
    \centering
    \includegraphics[width=\columnwidth]{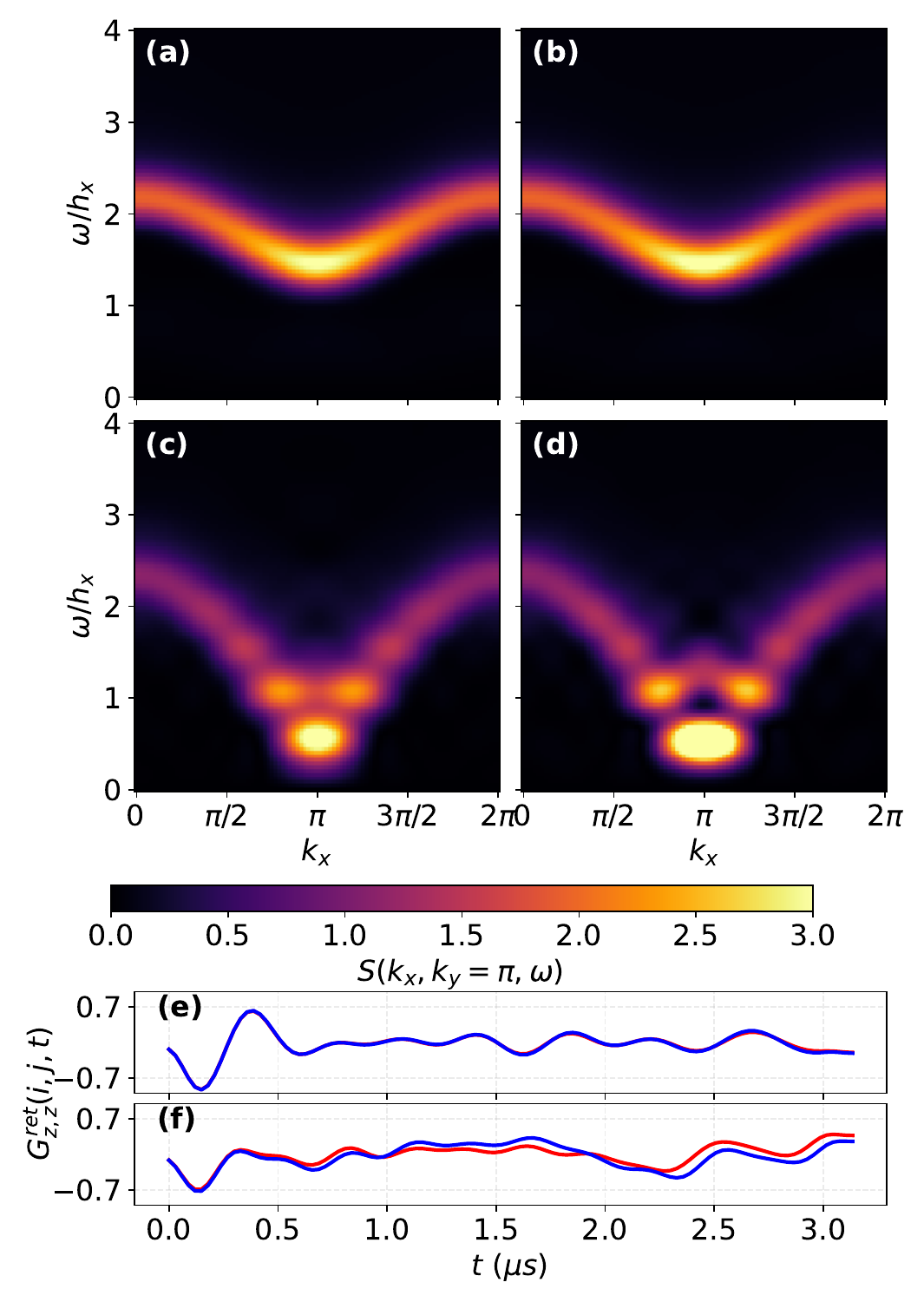}
    \caption{Bond-dimension dependence of the 2D dynamical structure factor ($6\times6$ TFIM) obtained via MPS emulation of a neutral-atom QPU protocol. (a-b) Band structure obtained at $J/h_x=0.2$ with bond dimensions $\chi=256$ (a) and $\chi=8$ (b). (c-d) Band structure obtained at $J/h_x=0.4$ with bond dimensions $\chi=256$ (c) and $\chi=8$ (d). (e-f) Time evolution of the retarded Green function of Eq.~(\ref{eq:model_retardedgreen}) (neglecting the third term) for a flipped qubit at the center of the lattice ($j=i$) for $J/h_x=0.2$ (e) and $J/h_x=0.4$ (f), showing the comparison between $\chi=8$ (blue) and $\chi=256$ (red).}
    \label{fig:bond_dimension}
\end{figure}

To emulate the $N=4\times 4=16$ square lattice, we used the exact state-vector emulator \texttt{emu-sv}. In the case of the $N=6\times 6=36$ square lattice we used the matrix-product-state emulator \texttt{emu-mps} with a maximum bond dimension of $\chi=512$ and an evolution time step $dt=5\,$ns. Since this finite bond dimension represents a truncation of the total Hilbert space, which is intractable with state-vector methods at this system size, we assess the robustness of the results against the choice of $\chi$. To this aim, we computed the dynamical structure factor of a two-dimensional $6\times6$ TFIM in the regime $J/h_x=0.4$ and $J/h_x=0.2$ using different bond dimensions. The corresponding results are reported in Fig.~\ref{fig:bond_dimension}.
In the paramagnetic regime considered in the main text ($J/h_x=0.2$) we observe that the DSF band is insensitive to changes in the bond dimension, as shown in Figs.~\ref{fig:bond_dimension}(a-b). This is further visualized in the time-resolved signals, shown in Fig.~\ref{fig:bond_dimension}(e). This convergence of the numerical methods is consistent with the fact that in this regime perturbative methods such as linear spin wave theory are expected to capture the main properties of the excitation band.
In contrast, we observe a clear dependence on the bond dimension closer to the parameter regime corresponding to a quantum phase transition in the thermodynamic limit, namely $J/h_x=0.4$. As can be seen in Figs.~\ref{fig:bond_dimension}(c-d) in this case the DSF signal shows the proximity to a gap closing, and the signal is sensitive to the chosen bond dimension, as also visualized in the time-resolved signals of Fig.~\ref{fig:bond_dimension}(f).

\section{Analytical bands of the 1D Ising and SSH models}
\label{app:dsf}
The dynamical structure factor for the Ising model (i.e. for a non-dimerized system) can be computed analytically. The energy dispersion $\omega(k)$ comes out to be:
\begin{equation}
    \omega(k)=2\sqrt{(h_x^2+J^2)(1-\sin(2\theta)\cos(k))}
\end{equation}
with the angle $\theta$ defined as:
\begin{equation}
    \theta=\arcsin(\pm\frac{h_x}{\sqrt{h_x^2+J^2}})
\end{equation}
where the sign in front of the fraction is $+$ for the ferromagnetic $(J<0)$ Ising model and $-$ for the antiferromagnetic case $(J>0)$. The analytical dispersion relation is obtained from the exact Jordan–Wigner and Bogoliubov solution of the one dimensional TFIM. In the thermodynamic limit, the excitation spectrum is independent of the choice of boundary conditions. For finite systems, periodic and open boundary conditions differ only in the quantization of the allowed momenta and in finite size corrections, while the functional form of the dispersion remains unchanged.
In the large field limit, $h_x\gg J$, we have:
\begin{align}
    \theta\rightarrow\frac{3}{2}\pi \\
    \omega(k)\rightarrow 2h_x.
\end{align}
The DSF $S^{ZZ}(k,\omega)$ is then expected to show a flat magnon band, i.e. constant in $k$, at energy $\omega=2h_x$. \newline In general, the DSF will show a dispersive band with the lowest peak in energy equal to $\omega_{min}=\frac{2(h_x-J)}{h_x}$ occurring at $k=\pi$ and the highest peak equal to $\omega_{max}=\frac{2(h_x+J)}{h_x}$ occurring at $k=0$. In particular, at the ground state phase transition, i.e. when $h_x=J$, zero energy excitations are allowed, since $\omega_{min}=0$, with momentum $k=\pi$. In principle, in the thermodynamic limit (TDL) we expect to observe a completely filled excitation band while, because of the finite size of the system, called $N$, the DSF shows a non-zero signal only in exactly $N$ spots $(k,\omega)$ belonging to the band. Nevertheless, in real case scenarios, we are not able to access that amount of resolution for the DSF. In fact, besides the imperfection of state preparation, the evolution time is constrained by decoherence to at most $4\mu s$ and the number of measurements taken during the sequence is limited to a few tens or hundreds. As a consequence, the DSF computed with the protocol described in  Sec.~\ref{subsec:model_DSF_Green} is expected to be way less resolved than the ideal one.

When the system is dimerized, the model can be mapped into the SSH model, whose energy band is computed starting from the fermionic picture. It is well-known that the dimerization splits the single particle excitation spectrum into two bands separated by a gap of size $2|J-J'|$, where $J$ is the intradimer interaction, i.e. the interaction between qubits belonging to the same dimer, while $J'$ is the interdimer interaction, i.e. the interaction between qubits belonging to adjacent dimers, weaker with respect to $J$. Since we set $J>J'$, the absolute value in the expression of the gap can be removed. The full expression for the excitation band reads:
\begin{equation}
\omega(k)=\pm\bigg(2h_x\pm\sqrt{J^2+J'^2\pm2JJ'\cos(k)}\bigg).
\end{equation}

In large dimerization and strong field regimes, the spectral weight is expected to concentrate around the inner band edges located at $\omega_{\pm}=2h_x\pm|J-J'|$ which are separated by the SSH gap $2|J-J'|$, as mentioned before, and the band is flat. When $J'$ is not negligible compared to $h_x$ and the dimerization $J/J'$ is of the order of a few units, the band is more dispersive and, compared to the standard Ising model, the minimum and maximum energy of the band $\omega_\pm$ are shifted by $\pm|J-J'|$.

\section{Details on Fourier Transform}
\label{app:FT}

In this appendix, we describe the numerical procedure used to obtain the dynamical structure factor (DSF) from the time-dependent magnetization measured after a local rotation. In the following, we focus on the $zz$ response relevant for the DSF.
Let $i$ denote the site at which the magnetization is measured and $j$ the site on which the local rotation is applied. For each value of the evolution time $t$, the simulation provides the magnetization
\begin{equation}
S(i,j,t)=\left\langle\sigma_i^z(t)\right\rangle_j.
\end{equation}
In addition, we compute the reference magnetization in the absence of the local rotation,
\begin{equation}
S_0(i,t)=\left\langle\sigma_i^z(t)\right\rangle_0 .
\end{equation}
Within the local-rotation protocol, the combination
\begin{equation}
G_{ij}(t)=S(i,j,t)-\frac{1}{2}S_0(i,t)
\label{eq:measured_response}
\end{equation}
provides the real space, time-dependent response entering the retarded Green's function (see main text for further details).

The spatial Fourier transform is defined as
\begin{equation}
G(k,t)=\frac{1}{L}\sum_{i,j}G_{ij}(t)e^{-ik r_{ij}} .
\label{eq:spatialFT}
\end{equation}

Eq. ~\eqref{eq:spatialFT} is evaluated directly for a prescribed set of momenta $k$, rather than using a standard discrete FT. This allows the momentum grid to be chosen independently of the system size.

The numerical protocol directly provides the response only for positive times, $t>0$. To perform the frequency-domain transform, we construct a two-sided time-domain function by imposing the Hermiticity relation
\begin{equation}
G(k,-t)=-G(k,t)^*,\qquad t>0.
\label{eq:negative_time_extension}
\end{equation}

The reason for eq.~\eqref{eq:negative_time_extension} can also be seen directly at the level of the Fourier transform. Neglecting for the moment the finite time window, define
\begin{equation}
G(k,\omega)=\int_{-T}^{T}dt\,G(k,t)e^{i\omega t}.
\label{eq:full_time_FT}
\end{equation}
Splitting the integral into positive and negative times gives
\begin{align}
G(k,\omega)=&\int_0^T dt\,G(k,t)e^{i\omega t}+\int_{-T}^0 dt\,G(k,t)e^{i\omega t}
\\
=&
\int_0^T dt\,G(k,t)e^{i\omega t}+
\int_0^T dt\,G(k,-t)e^{-i\omega t}.
\end{align}
Using eq.~\eqref{eq:negative_time_extension},
\begin{align}
G(k,\omega)&=\int_0^T dt\left[G(k,t)e^{i\omega t}-G(k,t)^*e^{-i\omega t}\right]
\\
&=2i\int_0^T dt\,\operatorname{Im}\left[G(k,t)e^{i\omega t}\right].
\label{eq:pure_imaginary_transform}
\end{align}
Therefore,
\begin{equation}
\operatorname{Re}G(k,\omega)=0
\end{equation}
and the resulting frequency-domain response is purely imaginary.

In the numerical implementation, we additionally multiply the time-domain data by a symmetric Parzen window $W(t)$ in order to reduce finite-time spectral leakage. Since the window satisfies
\begin{equation}
W(-t)=W(t),
\end{equation}
the same argument applies and eq.~\eqref{eq:pure_imaginary_transform} becomes
\begin{equation}
G(k,\omega)=2i\int_0^T dt,W(t)\,\operatorname{Im}\left[G(k,t)e^{i\omega t}\right].
\label{eq:windowed_transform}
\end{equation}
Hence the use of the window does not modify the purely imaginary character of the Fourier transform; it only affects the finite-time frequency resolution and suppresses spectral oscillations.

\end{document}